\documentclass[b5paper,twoside]{jpconf}  %Default declaration%
\usepackage{graphicx}

\begin{document}

\title[P-L Relations for Magellanic Clouds Cepheids]{Period-Luminosity Relations For Magellanic Clouds Cepheids Based on OGLE-III Data: A Comparison}  

\author[Ngeow]{Chow-Choong Ngeow$^1$}

\address{$^1$Graduate Institute of Astronomy, National Central University, Jhongli City, 32001, Taiwan}

\ead{cngeow@astro.ncu.edu.tw} 

\begin{abstract}

The period-luminosity (P-L) relation for Cepheid variables is important in modern astrophysics. In this work, we present the multi-band P-L relations derived from the Large Magellanic Cloud (LMC) and Small Magellanic Cloud (SMC) Cepheids, based on the latest release of OGLE-III catalogs. In addition to the $VI$ band mean magnitudes adopted from OGLE-III catalogs, we also cross-matched the LMC and SMC Cepheids to the 2MASS point source catalogs and publicly available {\it Spitzer} catalogs from SAGE program. Mean magnitudes for these Cepheids were corrected for extinction using available extinction maps. When comparing the P-L slopes, we found that the P-L slopes in these two galaxies are consistent with each others within $\sim2.5\sigma$ level.

\end{abstract}

\section{Introduction}

The period-luminosity (P-L, also known as Leavitt Law) relation for Cepheid variables is important in modern astrophysics, as it is the first rung in the distance scale ladder. The P-L relations can also be used to constrain the theoretical PL relations based on stellar pulsation and evolution models. One important issue in the application of P-L relation in distance scale work is its universality -- is the slope of P-L relation independent of metallicity? In this work, the multi-band P-L relations were derived for the Large Magellanic Cloud (LMC) and Small Magellanic Cloud (SMC) Cepheids, based on the latest release of OGLE-III (the third phase of Optical Gravitational Lensing Experiment) catalogs. Large numbers ($\sim 10^3$) of Cepheids in the LMC and SMC permit the determination of accurate P-L slopes, hence testing the universality of the P-L relation in low metallicity galaxies.

\section{Data and Methods}

Periods and intensity mean magnitudes in $VI$ bands were available from OGLE-III catalogs for $\sim1800$ LMC Cepheids \cite{sos08} and $\sim2600$ SMC Cepheids \cite{sos10}. All of these Cepheids are fundamental mode Cepheids. SMC Cepheids with $\log(P)<0.4$, however, were removed from the sample, as they followed different P-L relation \cite{bau99}. These Cepheids were cross-matched to 2MASS point source catalog. The random-phase $JHK$ photometry were converted to mean magnitudes using the prescription given in \cite{sos05}.  The mid-infrared photometry were available from {\it Spitzer} archival data -- SAGE \cite{mei06} and SAGE-SMC. Zaritsky's extinction maps for LMC \cite{zar04} and SMC \cite{zar02} were used for extinction corrections. In addition, the extinction-free P-L relation -- the Wesenheit Function in the form of $W=I-1.55(V-I)$, was also derived. Outliers presented in the P-L plane were removed using an iterative sigma clipping algorithm \cite{nge09}. Additional period cuts need to be applied in certain bands, as the faint end (hence shorter period) of these P-L relations may be affected by incompleteness bias.

\section{Results and Conclusion}

\begin{table}
\caption{\label{tab}Slopes of the P-L Relations.}
\begin{center}
\begin{tabular}{llll}
\br
Band & LMC P-L Slopes & SMC P-L Slopes & Slope Difference \\
\mr
$V$   & $-2.769\pm0.023$ & $-2.672\pm0.036$ & $ 0.097\pm0.043$ \\
$I$   & $-2.961\pm0.015$ & $-2.926\pm0.028$ & $ 0.035\pm0.032$ \\
$J$   & $-3.115\pm0.014$ & $-3.062\pm0.024$ & $ 0.053\pm0.028$ \\
$H$   & $-3.206\pm0.013$ & $-3.171\pm0.023$ & $ 0.035\pm0.026$ \\
$K$   & $-3.194\pm0.015$ & $-3.231\pm0.039$ & $ 0.037\pm0.038$ \\
$3.6\mu\mathrm{m}$ & $-3.253\pm0.010$ & $-3.226\pm0.019$ & $0.027\pm0.021$ \\
$4.5\mu\mathrm{m}$ & $-3.214\pm0.010$ & $-3.184\pm0.020$ & $0.030\pm0.022$ \\
$5.8\mu\mathrm{m}$ & $-3.182\pm0.020$ & $-3.235\pm0.042$ & $0.053\pm0.047$ \\
$8.0\mu\mathrm{m}$ & $-3.197\pm0.036$ & $-3.281\pm0.062$ & $0.084\pm0.072$ \\
$W$ & $-3.313\pm0.008$ & $-3.319\pm0.018$ & $ 0.006\pm0.020$ \\
\br
\end{tabular}
\end{center}
\end{table}

Comparison of the multi-band P-L slopes for these two metal poor galaxies are presented in Table \ref{tab}. Note that the LMC P-L slopes have been published in \cite{nge09}. As can be seen from this Table, the P-L slopes are within $\sim2.5\sigma$ in all bands between the LMC and SMC Cepheids. The $W$ band P-L slopes are almost identical, suggesting the extinction-free $W$ band P-L relation is a good choice for distance scale application.

\ack
CCN thanks the funding from National Science Council (of Taiwan) under the contract NSC 98-2112-M-008-013-MY3.

\end{document}